\documentstyle[aps,bezier,epsfig,twoside]{revtex}
\newcommand{\C}{{\if mm {{\rm C}\mkern -15mu{\phantom{\rm t}\vrule}}
\mkern +10mu \else \leavemode \hbox{I}\kern -.17em \hbox{C} \fi}}
\begin{document}
\thispagestyle{empty}
\title{Soliton Model of Hydrogen Atom: Resonance Effects}
\author{Yu.P.~Rybakov \\
{\small \it Department of Theoretical Physics}\\
{\small \it Russian Peoples' Friendship University}\\
{\small \it 6, Miklukho-Maklay str., 117198 Moscow, Russia\\
e-mail: Yurii.Rybakov@mx.pfu.edu.ru}}
\author{B.~Saha \\
{\small \it Bogoliubov Lab. of Theoretical Physics }\\
{\small \it Joint Institute for Nuclear Research}\\
{\small \it 141980 Dubna, Moscow Reg., Russia\\
e-mail: saha@thsun1.jinr.dubna.su}}
\maketitle
\begin{abstract}
Some first principles that, we believe, could serve as foundation
for quantum theory of extended particles are formulated.
 It is also shown that in the point-like particles limit
the non-relativistic quantum mechanics can be restored. As an illustration the soliton
model of hydrogen atom is considered.
\end{abstract}

To begin with we formulate the first principles for quantum theory of
extended particles:
\begin{itemize}
\item
Following A. Einstein and L. de Broglie we describe the extended particles
by the stable soliton-like solutions to non-linear field equations.
\item
Along the line of D. Bohm's thought we accept that the wave properties
of particles have the origin in non-linear resonance effect.
\item
We assume that the statistical properties of particles can be
deduced in the point-like limit from an analog of the wave function
describing the quantum statistical ensemble of D. Blokhintsev.
\end{itemize}
To illustrate these principles we consider the simplest scalar field model
given by the Lagrangian in the Minkowski space-time
\begin{eqnarray}
L_0 = \partial_i \phi^{*} \partial_j \phi \eta^{ij} - (mc/\hbar)^2
\phi^{*} \phi + F (S), \quad S = \phi^{*} \phi,
\label{lag}
\end{eqnarray}
with $F (S)$ behaving as $S^n$, $n > 1$, for $S \to 0$. This model admits,
for many choices of $F$, e.g., $F = k S^n,\,\, k> 0,\,\, 1< n < 5/3$, stable
soliton-like solution of stationary type
\begin{equation}
\phi_0 = u (r) e^{-i \omega_0 t}, \quad r = |{\bf r}|,
\label{sol}
\end{equation}
with the energy
\begin{equation}
E = \int\, d^3 x T^{00} (\phi_0)
\label{en}
\end{equation}
and the electric charge
\begin{equation}
Q = e \omega_0 \int\, d^3 x |u|^2.
\label{char}
\end{equation}
D. Bohm in his book "Causality and Chance in Modern Physics" (1957) discussed
the following problem. Let $\phi = \phi_0 + \xi (t, {\bf r})$ describes the
perturbed soliton-like solution. D. Bohm put the following question:
Does there exist any nonlinear model for which the spatial asymptote of
$\xi (r \to \infty)$ represents oscillations with characteristic frequency
\begin{equation}
\omega=E/\hbar?
\label{res}
\end{equation}
As is clear from the structure of the Lagrangian (\ref{lag}), at spatial
infinity the field equation reduces to the linear Klein-Gordon one
\begin{equation}
[\Box -(mc/\hbar)^2]\phi = 0,
\label{k-g}
\end{equation}
and therefore the principle of non-linear resonance by Bohm (\ref{res})
holds only for solitons with the energy $E = m c^2$. It that the
universality of the Planck-de Broglie relation (\ref{res}) fails.
To reinstate the universality of the relation (\ref{res}) we modify
the model (\ref{lag}) including gravity:
\begin{eqnarray}
L = c^4 R/16 \pi G + \partial_i \phi^{*} \partial_j \phi \eta^{ij}
- I (g_{ij}) \phi^{*} \phi + F (\phi^{*} \phi).
\label{modlag}
\end{eqnarray}
The crucial point of the model is to choose the invariant $I (g_{ij})$
with the asymptotic property
\begin{equation}
\lim\limits_{r \to \infty} I (g_{ij}) = (mc/\hbar)^2,
\label{lim}
\end{equation}
where $m$ stands for the Schwarzschild mass of the soliton. It can be verified
that the relation (\ref{lim}) holds if one chooses
\begin{equation}
I=(I_{1}^{4}/I_{2}^{3})c^6\hbar^{-2}G^{-2},
\label{inv}
\end{equation}
where
$ I_1=R_{ijkl}R^{ijkl}/48, \quad I_2=-R_{ijkl;n}R^{ijkl;n}/432.$
Estimating $R^{ijkl}$ at large distance one finds
$ I_1=G^2 m^2/(c^4 r^6), \quad I_2=G^2 m^2/(c^4 r^8).$
Thus we conclude that the principle of wave-particle duality has the
gravitational origin in our model [1].
Now let us construct the analog of the wave function. Suppose that the field
$\phi$ describes $n$ particles and has the form
\begin{equation}
\phi (t, {\bf r}) = \sum_{k = 1}^{n} \phi^{(k)} (t, {\bf r}),
\label{phik}
\end{equation}
where $${\rm supp}\, \phi^{(k)}\, \cap \,\,
{\rm supp}\, \phi^{(k')} = 0,\quad k\ne k', $$ and
the same for the conjugate momenta
$$\pi (t, {\bf r}) = \partial L/\partial \phi_t = \sum_{k = 1}^{n}
\pi^{(k)} (t, {\bf r}), \quad \phi_t = \partial \phi/\partial t.$$
Let us define the auxiliary functions
\begin{equation}
\varphi^{(k)} (t, {\bf r}) = \frac{1}{\sqrt{2}} (\nu_k
\phi^{(k)} + i \pi^{(k)}/\nu_k)
\label{aux}
\end{equation}
with the constants $\nu_k$ satisfying the normalization condition
\begin{equation}
\hbar = \int\, d^3 x |\phi^{(k)}|^2.
\label{nc}
\end{equation}
Now we define the analog of the wave function in the configurational space
$\{{\bf r}_1, \cdots {\bf r}_n\} \in \Re^{3n}$ as
\begin{equation}
\Psi_N (t, {\bf r}_1, \cdots {\bf r}_n) = (\hbar^n N)^{-1/2} \sum_{i = 1}^{N}
\prod_{k = 1}^{n} \varphi_i^{(k)} (t, {\bf r}_k),
\label{wfcs}
\end{equation}
where $N \gg 1$ stands for the number of trials (observations) and
$\varphi_{i}^{(k)}$ is the one-particle function (\ref{aux}) for the $i$ -th
trial. It can be shown [1] that the quantity
$$\rho_N = \frac{1}{(\triangle \vee)^n}\,
\int\limits_{(\triangle \vee)^n \subset{\Re^{3n}}} d^{3n}x |\Psi_N|^2, $$
where $\triangle \vee$ is the elementary volume which is supposed to be
much greater than the proper volume of the particle
$\vee_0 \ll \triangle \vee$, plays the role of coordinate probability
density.
If we choose the classical observable $A$ with the generator $\hat{M}_A$,
one can represent it in the form
\begin{equation}
A_j = \int\, d^3 x \pi_j i \hat{M}_A \phi_j = \sum_{k = 1}^{n} \int\,
d^3 x \varphi_{j}^{* (k)} \hat{M}_{A}^{(k)} \varphi_{j}^{(k)},\nonumber
\end{equation}
for the $j$ - th trial. The corresponding mean value is
\begin{eqnarray}
< A > &=& \frac{1}{N} \sum_{j = 1}^{N} A_j = \frac{1}{N} \sum_{j = 1}^{N}
\sum_{k = 1}^{n} \int\,
d^3 x \varphi_{j}^{* (k)} \hat{M}_{A}^{(k)} \varphi_{j}^{(k)} \nonumber \\
&=& \int\,d^3 x \Psi_{N}^{*} \hat{A} \Psi_{N} +
O (\frac{\vee_0}{\triangle \vee})
\label{mv}
\end{eqnarray}
where the hermitian operator $\hat{A}$ reads
\begin{equation}
\hat{A} = \sum_{k = 1}^{n} \hbar \hat{M}_{A}^{(k)}.
\label{ho}
\end{equation}
Thus, upto the terms of the order $\vee_0/\triangle \vee \ll 1$, we
obtain the standard quantum mechanical rule for the calculation of mean values [1].
It is interesting to underline that the solitonian scheme contains also
the well-known spin - statistic correlation [1]. Namely, if $\varphi_{i}^{(k)}$
is transformed under the group rotation by irreducible representation
$D^{(J)}$ of $SO (3)$, then the transposition of two identical extended
particles is equivalent to the relative $2\pi$ rotation of $\varphi_{i}^{(k)}$
that gives the multiplication factor $(-1)^{2J}$ in $\Psi_N$.
It can be also proved that $\Psi_N$ upto the terms of order
$\vee_0/\triangle \vee$ satisfies the standard Schr$\ddot o$dinger equation [1].
Now we apply the solitonian scheme to the hydrogen atom [2]. Let us introduce
the nucleus Coulomb field
$ A_{i}^{\rm ext} = \delta_{i}^{0} Z e/r$
and consider the scalar field Lagrangian density
\begin{equation}
{\cal L}=-\frac{1}{16\pi}(F_{ik})^2 + |[\partial_k - i\varepsilon(A_k +
A_{k}^{\rm ext})]\phi|^2 - (mc/\hbar)^2\phi^*\phi + F(\phi^* \phi),
\label{sflag}
\end{equation}
where $\varepsilon = e / \hbar c$. Suppose that for $A_{k}^{\rm ext} = 0$
the field equations admit stable stationary soliton-like solution of type
(\ref{sol}) describing configurations with mass $m$ and electric charge
$e$. For simplicity we omit the gravitational field supposing that it has
been taken into account due to the non-linear resonance condition (\ref{res}).
Then, in the non-relativistic approximation we may put
\begin{equation}
\phi = \psi\,  \exp{(-imc^2t/\hbar)}.
\label{nra}
\end{equation}
Therefore, the corresponding field equations read
\begin{eqnarray}
& &i\hbar\, \partial_t \psi + (\hbar^2/2m)\triangle \psi + (Ze^2/r)\psi\,
= -(\hbar^2/2m) {\hat f}({\bf A}, A_0, \psi^*\psi)\psi \nonumber\\
& &\quad\equiv -(\hbar^2/2m) \Bigl[2i\varepsilon({\bf A} \nabla)\psi + 2(\varepsilon
mc/\hbar) A_0 \psi +i\varepsilon \psi\, \mbox{div} {\bf A} +
F^{\prime}(\psi^* \psi)\psi\Bigr],  \label{nls}\\
& &\Box A_0 =  (8\pi me/\hbar^2)|\psi|^2 \equiv -4\pi \varrho, \label{m0}\\
& &\Box {\bf A} = 4\pi [2\varepsilon^2
{\bf A} |\psi|^2 - i\varepsilon(\psi^* \nabla \psi - \psi \nabla \psi^*)]
\equiv -(4\pi /c)\, {\bf j}, \label{m3}\\
& &\partial_t A_0 + c\, {\rm div} {\bf A} = 0 \label{lc}
\end{eqnarray}
We will seek for the solutions to these equations
describing a stationary state of an atom when the electron - soliton
center moves along a circular orbit of radius $a_0$ with some angular
velocity $\Omega$. We have two characteristic lengths in this problem:
the size of the soliton $\ell_0 = \hbar/mc$ and the Bohr radius
$a = \hbar^2/mZe^2 \gg \ell_0$. Near the soliton center, where
$r - a_0 \le \ell_0$, we get in non-relativistic approximation
\begin{eqnarray}
\psi &=& u ({\bf R}) e^{iS/\hbar} = \psi_{-}, \quad
A_0 = A_0 ({\bf R}), \quad {\bf A} = \frac{1}{c} \dot{\bf \xi} (t)
A_0 ({\bf R}) \nonumber
\end{eqnarray}
with
\begin{eqnarray}
S &\approx& m \dot{\bf \xi} \cdot {\bf R} + C_0 t + \chi (t), \quad
m \ddot{\bf \xi} = - Z e^2 {\bf \xi}/\xi^3, \nonumber\\
\chi (t) &=& \int\limits_{0}^{t} \bigl(\frac{m}{2} \dot{\bf \xi}^2
+ \frac{Z e^2}{\xi}\bigr) dt - \quad {\rm the\,\,\, Hamiltonian \,\,\, action.}
\nonumber
\end{eqnarray}
The function $u (\bf R)$, where ${\bf R} = {\bf r} - {\bf \xi}(t)$ satisfies
the following soliton-like equation
$\hbar^2 (\hat f + \triangle u /u)= 2 m C_0.$
For $\psi$ we have the integral equation
\begin{eqnarray}
\label{int}
\psi (t, {\bf r}) &=& C_n \psi_n({\bf r})\, \exp{(-i\omega_nt)} \\
&+& \frac{1}{2\pi}\int d\omega \int dt^{\prime} \int d^3x^{\prime}\,
\exp{[-i\omega(t-t^{\prime})]}\,G({\bf r}, {\bf r}^{\prime};\omega+i0)
\hat f \psi (t^{\prime}, {\bf r}^{\prime}), \nonumber
\end{eqnarray}
with $G$ being the Coulomb resolvent, $E_n = \hbar \omega_n$ is the
eigenvalue of the Coulomb Hamiltonian. For $R \gg \ell_0$ we  may put
in (\ref{int})
$$\hat f\, \psi (t, {\bf r})= g\, \exp{(-i\omega_n t)}\,\delta({\bf r} -
{\bf \xi}(t)), \quad g=const.$$
Calculating the integral (\ref{int}) by stationary phase method we get
\begin{eqnarray}
\psi = \psi _{+} \approx C_n \psi_n({\bf r})\, e^{-i\omega_n t}
- \frac{g |\omega_n| m a}{8\pi^2 \hbar \sqrt{a_0 {\rm cos}^2(\vartheta/2)}}
e^{-i\omega_n t} R^{-3/2} e^{-R\sqrt{2 m |\omega_n|/\hbar}}, \nonumber
\end{eqnarray}
where ${\rm cos} \vartheta = {\rm sin} \theta {\rm cos} (\alpha - \Omega t)$.
Now to find the constants $C_0, C_n, a_0, \Omega, g$ we must match the
functions $\psi_{+}$ and $\psi_{-}$ at $R = \ell_0$. That gives the following
results
\begin{eqnarray}
a_0 &=& a n,\quad \Omega^2 = Z e^2/m a_0^3, \quad C_0 = - m \Omega^2 a_0^2,
\nonumber\\
C_n \psi_n (a_0) &=&
 \frac{g |\omega_n| m a}{8\pi^2 \hbar \sqrt{a_0}}
\ell_0^{-3/2} e^{-\ell_0\sqrt{2 m |\omega_n|/\hbar}} + u (\ell_0), \nonumber\\
g &=& \int\limits_{\vee_0} d^3 x \hat{f} u, \quad
\vee_0 = \frac{4}{3} \pi \ell_{0}^{3}. \nonumber
\end{eqnarray}
The last step is the calculation of the electromagnetic field for
$R \gg \ell_0$ and for large time $t \gg 1/|\omega_n|$, that gives the
semi-sum of the retarded and advanced potentials:
$A_\mu = \frac{1}{2} \bigl(A_{\mu}^{\rm adv} + A_{\mu}^{\rm ret}\bigr)$.
It is interesting to write down the components of the Poynting vector
$\bf S$:
\begin{eqnarray}
S_r&=&\frac{e^2\,a_{0}^{2}\,\Omega^4}{16 \pi c^3\,r^2}\,
\mbox{sin}^2\vartheta\, \mbox{sin} 2(\alpha - \Omega
t)\, \mbox{sin} (2\Omega r/c),  \nonumber \\
S_\vartheta &=& \frac{e^2\,a_0\,\Omega^2}{4 \pi c\,r^3}\,
\mbox{cos}\vartheta\, \mbox{sin} (\alpha - \Omega
t)\, \mbox{sin} (\Omega r/c),  \nonumber \\
S_\alpha &=& \frac{e^2\,a_0\,\Omega^2}{4 \pi c\,r^3}\,
\,\mbox{cos} (\alpha - \Omega
t)\, \mbox{sin} (\Omega r/c). \nonumber
\end{eqnarray}
Thus we conclude that the radiation is absent. The various aspects of the
solitonian scheme were discussed in details in ~\cite{Rybakov,Saha}.

\end{document}